\documentclass[1p,times]{elsarticle}
\usepackage[T1]{fontenc} % Use modern font encodings    
\usepackage{multicol}
\usepackage{amsmath, amssymb}
\usepackage{setspace}
\usepackage{newunicodechar}
\usepackage{epsfig}
\usepackage{graphicx}
\usepackage{gensymb}
\usepackage{natbib}
\usepackage{adjustbox}
\newunicodechar{≤}{\ensuremath{\leq}}
\usepackage{caption}
\usepackage{subcaption}
\usepackage{lipsum}% for dummy tex
\usepackage{hyperref}   % had to loaded last
\usepackage{float}
\usepackage{multirow}
\usepackage{multicol}
\usepackage{array}
\usepackage{xcolor,soul}
\sethlcolor{white}

\usepackage[labelfont={bf}]{caption}

\DeclareUnicodeCharacter{2212}{\ensuremath{-}}
\DeclareUnicodeCharacter{22C5}{\ensuremath{.}}
\DeclareUnicodeCharacter{2009}{\ensuremath{:}}
\newcommand\redsout{\bgroup\markoverwith{\textcolor{yellow}{\rule[0.5ex]{2pt}{1.5pt}}}\ULon}
\journal{Journal of Molecular Liquids}
\begin{document}

\begin{frontmatter}



%\title{Ion Pairing of ${\rm Zn^{2+}}-{\rm {SO_{4}}^{-}}$ and ${\rm Mg^{2+}}-{\rm {SO_{4}}^{-}}$ in Aqueous ${\rm Mg^{2+}}-{\rm {SO_{4}}^{-}}$/${\rm Zn^{2+}}-{\rm {SO_{4}}^{-}}$ Hybrid Electrolytes: Insights from Molecular Dynamics Simulations} 

\title{The Association of Zn\(^{2+}\)-SO\(_4^{2-}\) and Mg\(^{2+}\)-SO\(_4^{2-}\) in Aqueous MgSO\(_4\)/ZnSO\(_4\) Hybrid Electrolytes: Insights from All-Atom Molecular Dynamics Simulations}

%% Article title


%\author{} %% Author name

%%\affiliation{organization={},%Department and Organization
 %%           city={},
 %%           postcode={}, 
   %%         state={},
   %%         country={}}
            
\author[1]{Mayank Dixit\corref{cor1}}
\ead{dixit@cheme.kyoto-u.ac.jp
} %% Author name
\author[2]{Timir Hajari\corref{cor2}}
\ead{timir230@gmail.com}
\author[3]{Bhalachandra Laxmanrao Tembe\corref{cor3}}
\ead{bltembe@iitdh.ac.in}
\cortext[cor1]{Corresponding author}
\cortext[cor2]{Corresponding author}
\cortext[cor3]{Corresponding author}
\affiliation[1]{organization={Kyoto University},
addressline={Graduate School of Engineering},
postcode={615-8510},
city={Kyoto},
country={Japan}}

\affiliation[2]{organization={Raja Rammohan Sarani},
addressline={Department of Chemistry, City College, 102/1},
city={Kolkata},
postcode={700009},
country={India}}

\affiliation[3]{organization={Indian Institute of Technology Dharwad},
addressline={Department of Chemistry, Indian Institute of Technology Dharwad,
Dharwad, Karnataka, India - 580011},
city={Dharwad},
postcode={580011},
country={India}}

%% Abstract
\begin{abstract}
${\rm Mg}{\rm {SO_{4}}}$
 is utilized as an additive to mitigate capacity fading in rechargeable zinc-ion batteries.
In this study, we conducted an in-depth investigation into the association and solvation structure of ${\rm {Zn}^{2+}-{SO_{4}^{2-}}}$ and ${\rm {Mg}^{2+}-{SO_{4}^{2-}}}$ ion pairs in various mixtures of ${\rm {[{ZnSO_{4}}]_{2M}+[{MgSO{4}}]_{0M}}}$, ${\rm {[{ZnSO{4}}]_{1M}+[{MgSO{4}}]_{1M}}}$, and ${\rm {[{ZnSO{4}}]_{0M}+[{MgSO{4}}]_{2M}}}$. 
To achieve this, we employed all-atom molecular dynamics (MD) simulations to analyze the dynamics of these mixtures through the dipole-dipole autocorrelation function ${\phi}(t)$ and dipole relaxation time ${{\tau}_{\rm d}}$. We explored the spatial distributions of ${\rm Zn^{2+}}$ and ${\rm Mg^{2+}}$ around each other, as well as ${\rm SO_{4}^{2-}}$ and ${\rm H_{2}O}$, utilizing radial distribution functions (RDFs) and Running Coordination numbers. Additionally, we assessed the potentials of mean force (PMF) for the ion pairs and computed the preferential binding coefficients for ${\rm Zn^{2+}}$ and ${\rm Mg^{2+}}$ in the aforementioned mixtures.
Our findings reveal that transitioning from a solution containing ${\rm 2M {[{MgSO_{4}}]}}$ to an equimolar mixture of ${\rm 1M {[{ZnSO_{4}}]}}$ and ${\rm 1M {[{MgSO_{4}}]}}$ significantly reduces the association of ${\rm Zn^{2+}}$ ions, indicating a disruption in their self-association due to ${\rm Mg^{2+}}$ presence. Conversely, ${\rm Mg^{2+}}$ ions exhibit a slight increase in association, suggesting that ${\rm Mg^{2+}}$ ions have a greater tendency to associate in mixed salt environments. The RDFs show well-defined peaks around ${\rm Zn^{2+}}$ and ${\rm Mg^{2+}}$ in different mixtures, revealing differences in solvation structure and coordination environment. Notably, ${\rm Zn^{2+}}$ maintains a more ordered solvation structure with ${\rm SO_{4}^{2-}}$ ions compared to ${\rm Mg^{2+}}$, especially as ${\rm MgSO_{4}}$ concentration increases.
The PMF analysis indicates distinct structural motifs for both ion pairs, with ${\rm Zn^{2+}}$ exhibiting lower energy barriers for stabilization compared to ${\rm Mg^{2+}}$. Furthermore, the preferential binding coefficients suggest that both cations are preferentially solvated by ${\rm SO_{4}^{2-}}$, with ${\rm Zn^{2+}}$ showing an increased affinity for ${\rm SO_{4}^{2-}}$ as the concentration of ${\rm MgSO_{4}}$ varies. These results elucidate the nuanced interplay between ion pairing, solvation structure, and preferential binding in mixed salt systems, with implications for understanding the dynamics of ion interactions in various electrolyte environments.
\end{abstract}

\begin{graphicalabstract}
\centering
\includegraphics[height=0.2\textheight]{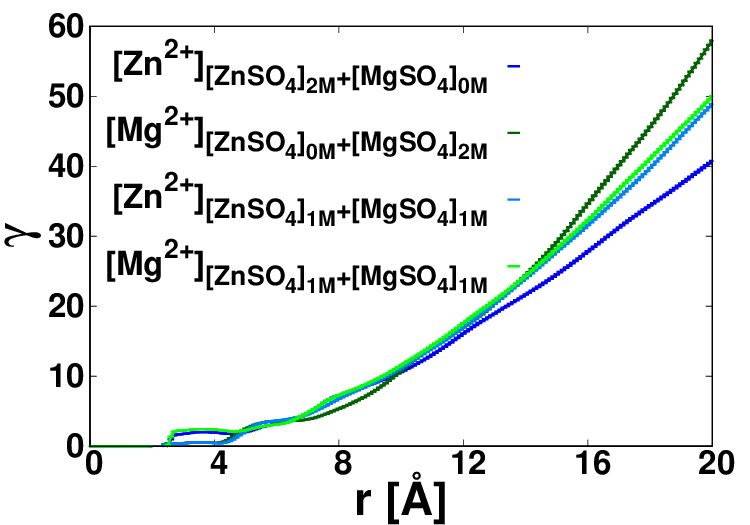}
\includegraphics[height=0.2\textheight]{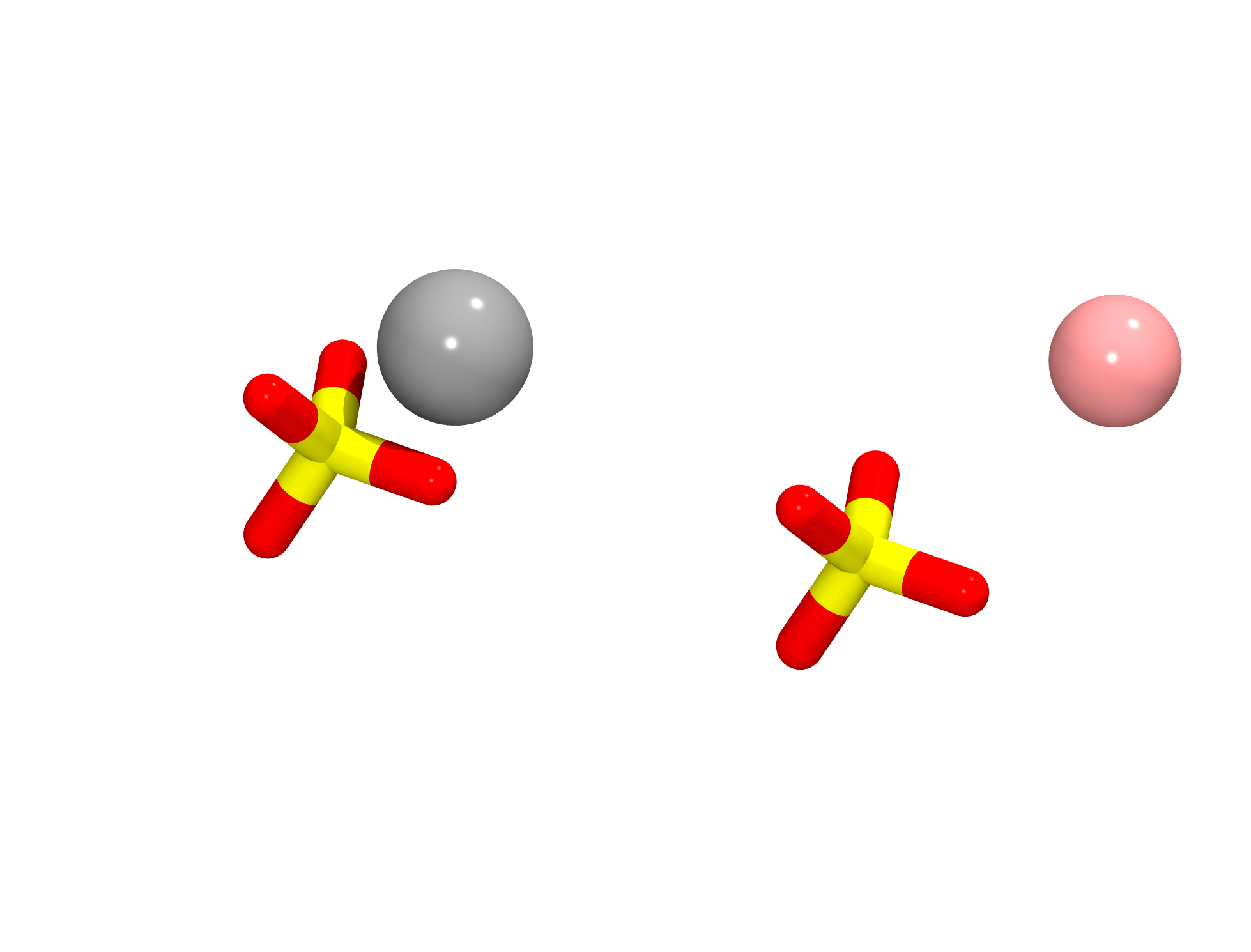}
\end{graphicalabstract}

\begin{highlights}
\item The addition of ${\rm MgSO_{4}}$ disrupts ${\rm Zn^{2+}}$ ion self-association, while ${\rm Mg^{2+}}$ ion association slightly increases.
\item In the ${\rm ZnSO_{4}}$ rich system, ${\rm Zn^{2+}}-{\rm SO_{4}^{2-}}$ ion pairs exhibit well-structured solvation compared to ${\rm Mg^{2+}}-{\rm SO_{4}^{2-}}$ pairs in ${\rm MgSO_{4}}$ rich systems.
\item The solvation environment around ${\rm Zn^{2+}}$ ions is more ordered and compact in mixed salt systems, with ${\rm Zn^{2+}}$ ions showing a higher propensity for self-association.
\item The addition of ${\rm MgSO_{4}}$ reduces the compactness of the solvation structure around ${\rm Zn^{2+}}$, while the solvation structure around ${\rm Mg^{2+}}$ becomes more ordered with higher ${\rm ZnSO_{4}}$ concentrations.
\item ${\rm Zn^{2+}}$ pairs form distinct contact and solvent-shared ion pairs with lower energetic barriers compared to ${\rm Mg^{2+}}$ pairs.
\item Both ${\rm Zn^{2+}}$ and ${\rm Mg^{2+}}$ are preferentially solvated by ${\rm SO_{4}^{2-}}$, with ${\rm Zn^{2+}}$ showing a stronger affinity in equimolar mixtures.
\end{highlights}

%% Keywords

\end{frontmatter}



\section{Introduction}
The increasing demand for grid-scale energy storage systems has driven the development of rechargeable battery technologies that prioritize low cost, eco-friendliness, and high operational safety\cite{Parker2017,Perez2017,Zhang2018,Xu2012,Liang2017,Huang2019,Xia2018,Kundu2018}. Rechargeable aqueous batteries, utilizing earth-abundant elements such as Na, K, Al, Mg, Ca, and Zn, present a promising alternative to non-aqueous systems for stationary grid-scale applications. These aqueous systems offer safer water-based electrolytes, higher ionic conductivity, and lower costs\cite{Parker2017,Perez2017,Zhang2018,Xu2012,Liang2017,Huang2019,Kundu2016,Ming2021,Ming2018,Zhang2017,Pan2016,Tang2018,Shan2019,Hu2018,Li2016,Konarov2018,Fang2018,Wu2018} Among these, zinc-ion batteries (ZIBs), which rely on $Zn^{2+}$ chemistry and a two-electron transfer mechanism, are gaining attention\cite{Roex2023,Li2020,Du2020,Hu2024,Zhang2023,Roex2023,Cang2020}. ZIBs utilize zinc metal as the anode material, offering high capacity in mildly acidic or neutral aqueous electrolytes\cite{Shi2018,Zhang2016,Huang2019}.
Despite the progress in ZIB development, the fundamental understanding of Zn-ion storage mechanisms remains limited. Electrolytes play a crucial role in forming protective layers on electrode surfaces and influencing the formation of byproducts like ZnO, $Zn(OH))_{2}$, or basic zinc sulfate\cite{Wan2018,Konarov2018,Zhou2018}. 
Therefore, the careful selection and preparation of appropriate electrolytes are just as critical as finding suitable cathode materials for the long-term stability and performance of ZIBs.
Vanadium- and manganese-based electrode materials often suffer from capacity fading during cycling, largely due to the dissolution of active materials into the electrolyte\cite{Li2016,Konarov2018}.
To mitigate this, researchers have found that pre-adding specific metal ions to the electrolyte can help stabilize the system. For instance, Oh et al. showed that the addition of ${\rm MnSO_{4}}$ improved the reversibility of the cathodic reaction and suppressed the formation of basic zinc sulfate on the ${\rm MnO_{2}}$ electrode\cite{Kim1998}. 
More recent studies confirmed the long cycling stability of ${\rm {MnO_{2}}}$ when manganese(II) salts like ${\rm MnSO_{4}}$ or ${\rm Mn(CF_{3}SO_{3})_{2}}$ were used as electrolyte additives\cite{Zhang2017,Pan2016}.
Similarly, metal vanadates experience dissolution of vanadium species (e.g., ${\rm VO^{2+}}$) during cycling, resulting in capacity decay\cite{Livage1991}.
The insertion of metal ions such as ${\rm Na^{+}}$, ${\rm K^{+}}$, ${\rm Zn^{2+}}$, ${\rm Ca^{2+}}$, or ${\rm Mg^{2+}}$ between${\rm V_{x}O_{y}}$ layers can act as stabilizing pillars. However, as these ions dissolve, the electrolyte may turn yellow due to the presence of free vanadium species. Recent research by Chen et al. demonstrated that the addition of $Na_{2}SO_{4}$ into a ${\rm Zn}{\rm {SO_{4}}}$ electrolyte can reduce vanadium dissolution, improving the stability of ${\rm Na{V_{3}}O_{8}}{.}{1.5H_{2}O}$\cite{Wan2018}. 
Recent research has focused on the investigation of electrolytes with varying concentration ratios of ${\rm Zn}{\rm {SO_{4}}}$ and ${\rm Mg}{\rm {SO_{4}}}$ for aqueous zinc-ion batteries.\cite{Zhang2020} Notably, batteries tested in a 1 M ${\rm Zn}{\rm {SO_{4}}}$−1 M ${\rm Mg}{\rm {SO_{4}}}$ electrolyte exhibited superior performance compared to other formulations, delivering a remarkable specific capacity of 374 mAh ${\rm g^{-1}}$ at a current density of 100 mA ${\rm g^{-1}}$. 
These batteries also demonstrated excellent rate performance with a stable reversible capacity. This study offers a promising approach to enhancing the performance of vanadium-based cathodes through electrolyte optimization using cost-effective solutions. However, the underlying molecular mechanism by which ${\rm Mg}{\rm {SO_{4}}}$ contributes to the performance improvements in zinc-ion batteries remains to be fully elucidated. The detailed mechanisms behind these stabilizing effects require further theoretical exploration.\\
In this study, we investigate the association and solvation structure of {${\rm Zn^{2+}}-{\rm {SO_{4}}^{2-}}$}  and ${\rm Mg^{2+}}-{\rm {SO_{4}}^{2-}}$  ion pairs in three different electrolyte solutions with varying concentration ratios of ${\rm Zn}{\rm {SO_{4}}}$ and ${\rm Mg}{\rm {SO_{4}}}$: 2.0 M ${\rm Zn}{\rm {SO_{4}}}$, 1.0 M ${\rm Zn}{\rm {SO_{4}}}$–1.0 M ${\rm Mg}{\rm {SO_{4}}}$, and 2.0 M ${\rm Mg}{\rm {SO_{4}}}$. These molecular insights are crucial for enhancing the electrochemical performance of aqueous ZIBs.
 Comprehensive details of our methodology and computational approach are expounded are given in section 2. Our results are comprehensively presented in section 3, ultimately culminating in our discussions and concluding remarks in section 4.

\begin{table*}[htbp]
\centering
%\resizebox{\columnwidth}{!}{%
\begin{adjustbox}{width=1.0\columnwidth,center}
\begin{tabular}{|l|c|c|c|c|c|c|c|c|}
\hline 
{System Code} &  ${n_{\rm Mg^{2+}}}$ &${n_{\rm SO_{4}{^{2-}}}}$ & ${n_{\rm Zn^{2+}}}$ & ${n_{\rm Water}}$ & ${\rho}({\rm {kg/m^3}})$ & {NVT-EQ} &  {NPT-EQ} & {NPT production run} \\ 
\hline 
$[{\rm ZnSO_{4}}]_{\rm 2M}$~~ (${\rm I}$) & 0  &  620 & 620 &15000&  1334.65$\pm$1.1 & 20ns &100ns & 2500ns \\ 
\hline 
$[{\rm ZnSO_{4}}]_{\rm 1M}$+$[{\rm MgSO_{4}}]_{\rm 1M}$ (${\rm II}$)  & 1  & 620 & 620 &15000&  1295.53$\pm$0.35  & 20ns & 100ns & 2500ns \\ 
\hline 
$[{\rm MgSO_{4}}]_{\rm 2M}$ (${\rm III}$) & 620 & 0  & 620 &15000&   1267.4$\pm$0.062 & 20ns & 100ns & 2500ns \\ 
\hline 
\end{tabular} 
\end{adjustbox}
\caption{${n_{\rm Mg^{2+}}}$ =  Number of ${\rm Mg^{2+}}$ ions; ${n_{\rm SO_{4}{^{2-}}}}$ = Number of ${\rm SO_{4}{^{2-}}}$ ions; ${n_{\rm Zn^{2+}}}$ =Number of ${\rm Zn^{2+}}$ ions;${n_{\rm Water}}$ =  Number of ${\rm Water}$ molecules  in the cubic simulation cell; $\rho$ = density at 298 $K$. The full-atom MD simulations were performed in the sequence i.e. NVT-EQ$\rightarrow$NPT-EQ$\rightarrow$NPT-production-run.}
\label{table:table_System}
\end{table*}

\begin{figure*}[hbtp]
\centering
\begin{subfigure}[b]{0.48\linewidth}
\centering
\caption{ ${\rm {[{ZnSO_{4}}]_{2M}+[{MgSO_{4}}]_{0M}}}$  }
\includegraphics[scale=0.09]{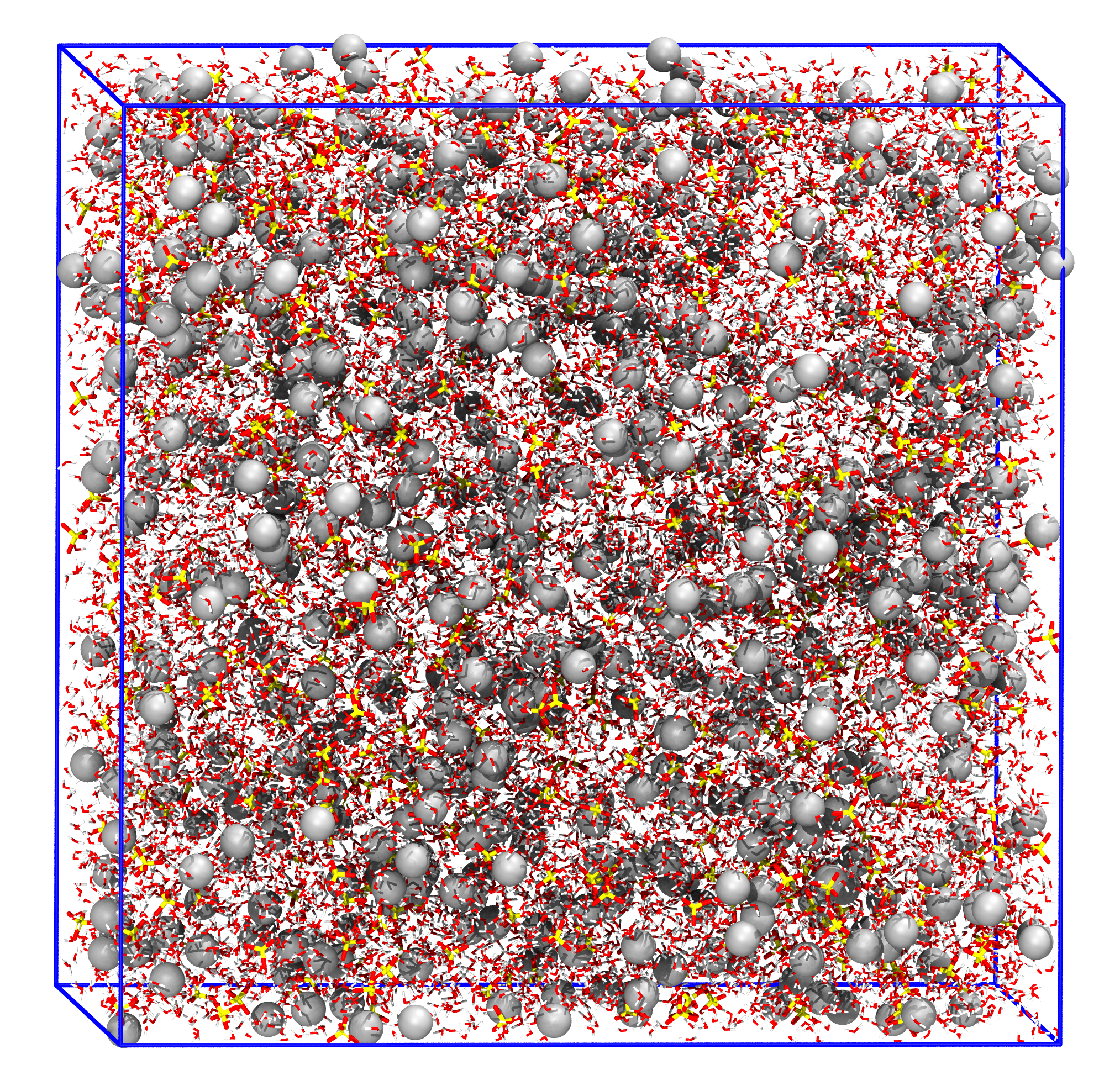}
\end{subfigure}
\begin{subfigure}[b]{0.48\linewidth}
\centering
\caption{ ${\rm {[{ZnSO_{4}}]_{1M}+[{MgSO_{4}}]_{1M}}}$  }
\includegraphics[scale=0.09]{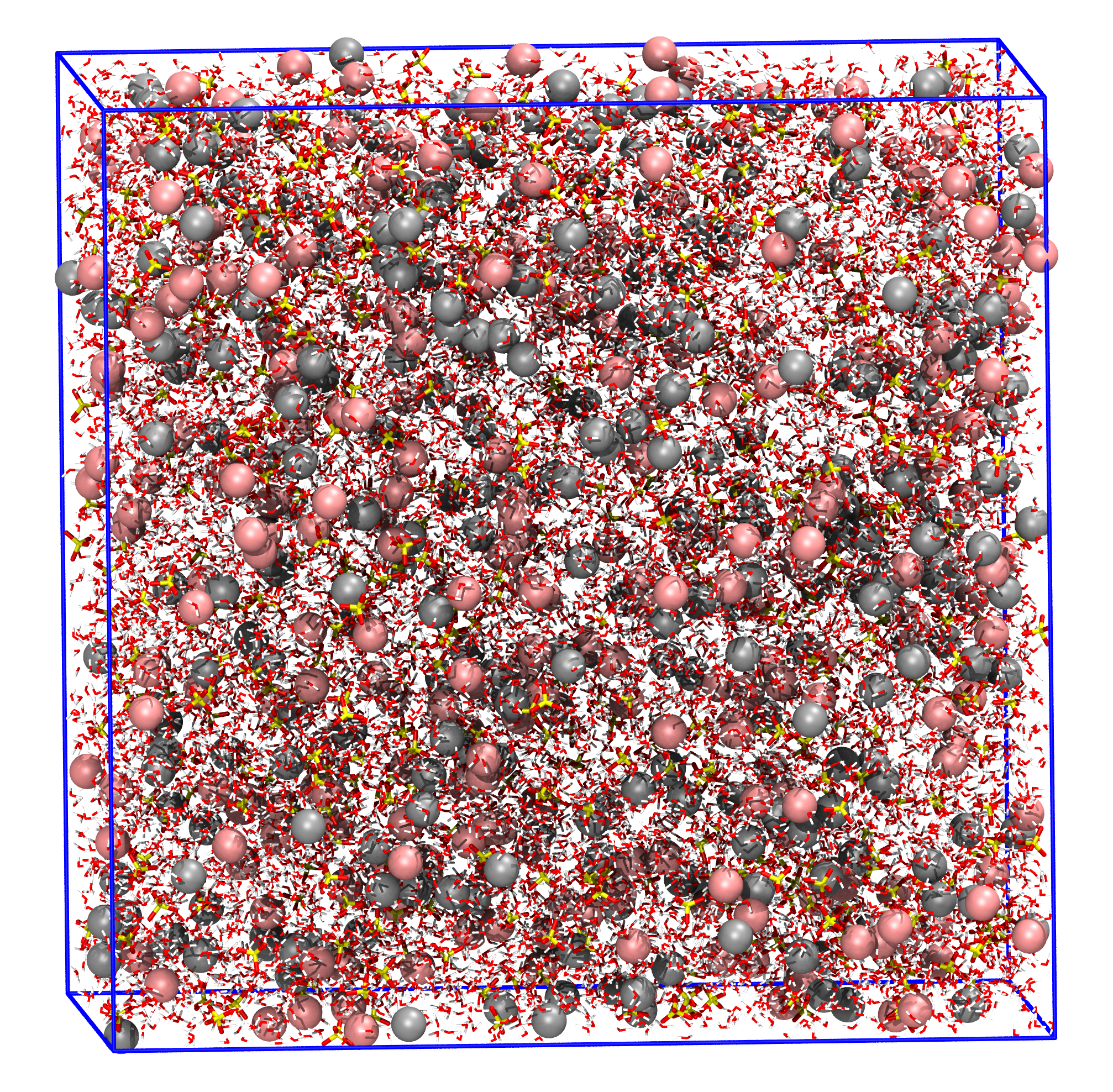}
\end{subfigure}

\begin{subfigure}[b]{1\linewidth}
\centering
\caption{ ${\rm {[{ZnSO_{4}}]_{0M}+[{MgSO_{4}}]_{2M}}}$  }
\includegraphics[scale=0.09]{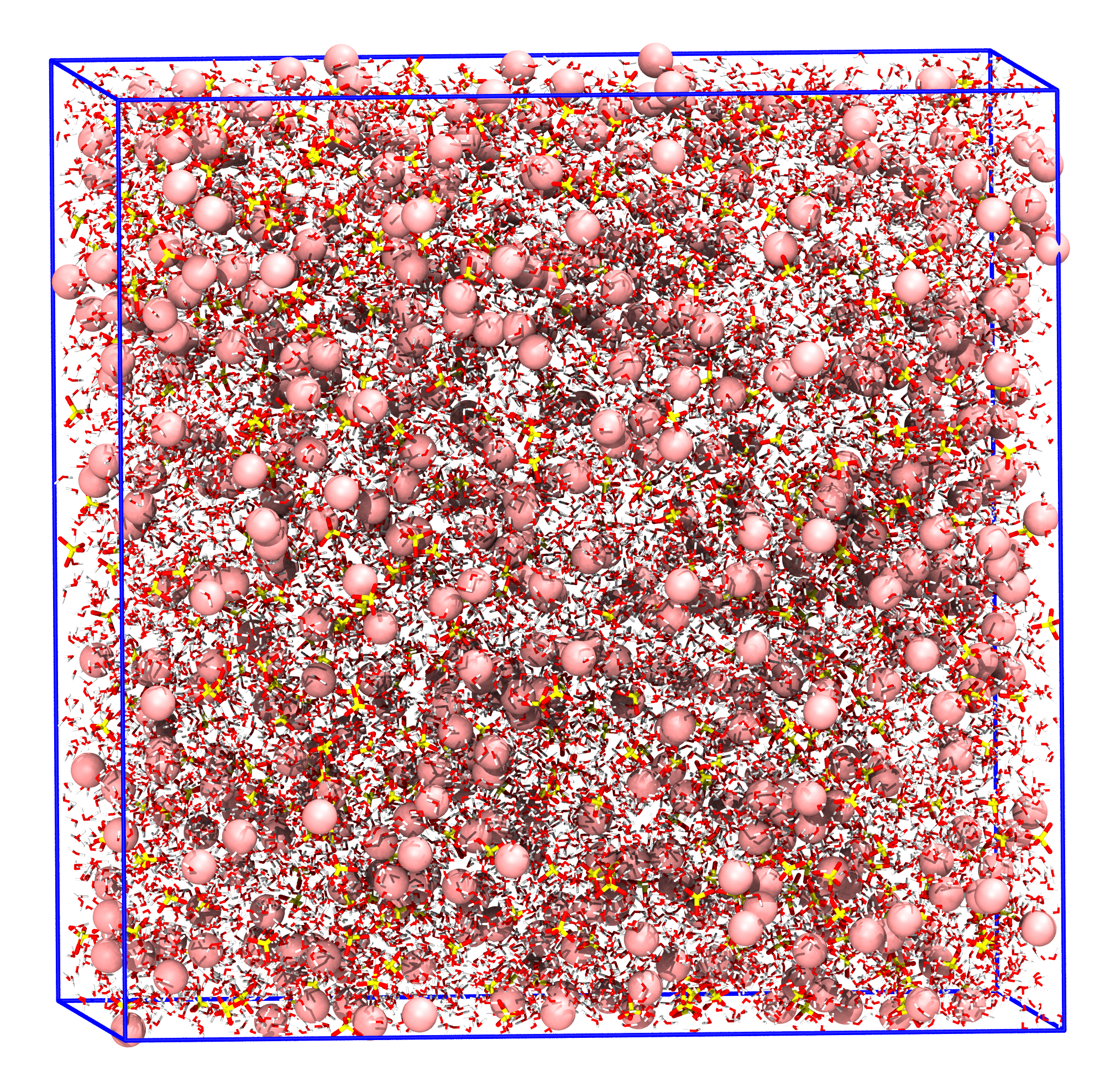}
\end{subfigure}

\caption{The initial configurations of ${\rm {[{ZnSO_{4}}]_{2M}+[{MgSO_{4}}]_{0M}}}$ (a), ${\rm {[{ZnSO_{4}}]_{1M}+[{MgSO_{4}}]_{1M}}}$ (b), ${\rm {[{ZnSO_{4}}]_{0M}+[{MgSO_{4}}]_{2M}}}$ (c). The Magnesium, sulpur, oxygen, Zinc, and hydrogen atoms are shown by pink, yellow, red, gray, white colors respectively.}
\label{Figure_Initial}
\end{figure*}

\newpage
\section{Methods and computational details}
In this study, we conducted an examination of ${\rm {[{ZnSO_{4}}]_{2M}+[{MgSO_{4}}]_{0M}}}$, ${\rm {[{ZnSO_{4}}]_{1M}+[{MgSO_{4}}]_{1M}}}$, ${\rm {[{ZnSO_{4}}]_{0M}+[{MgSO_{4}}]_{2M}}}$ mixtures (as outlined in Table \ref{table:table_System}). The initial structure of these mixtures are depicted in Figure \ref{Figure_Initial}.
\subsection{All-atom MD simulation details}
The initial configurations of the system were generated by packmol software\cite{Martinez2009} and the multicomponent assembler of CHARMMM-GUI.\cite{Fias2008} These initial configurations were subsequently employed for all-atom molecular dynamics (MD) simulations using GROMACS (version$-$2019.2)\cite{Abraham2015}. 
To generate the necessary input files compatible with GROMACS for each melt system, CHARMM-GUI was employed\cite{Brooks2009abcd,Jo2014,Lee2016abc}. The CHARMM general all-atom force field\cite{Won2012,Choi2022,Allouche2012,Manuscript2010}, was employed for the all$-$atom MD simulations. The TIP3P model was used for water molecules.\cite{Mark2001} The temperature and pressure were set to $298 {\rm K}$ and $1 {\rm bar}$, respectively.
In order to attain equilibrium state for each system outlined in Table \ref{table:table_System}, the initial configurations underwent an energy minimization process.
Subsequently, the energy-minimized structures were employed to perform NVT and NPT equilibrations (the simulation times for each step are presented in Table 1). The resulting equilibrated structures were then subjected to a 2500 ns production run. The generated trajectories were utilized to compute the dipole-dipole autocorrelation function, radial distribution function (RDF), and potentials of mean force $({\rm W{r}})$, preferential binding coefficients (${\gamma}$) for ${\rm {{{Zn}^{2+}}}}$, ${\rm {{{Mg}^{2+}}}}$, ${\rm {{SO_{4}}^{2-}}}$, and ${\rm {{H_{2}O}}}$ in ${\rm {[{ZnSO_{4}}]_{0M,1M,2M}+[{MgSO_{4}}]_{0M,1M,2M}}}$ mixtures. 
To control the system temperature, the Nos\'{e}-Hoover thermostat\cite{Nose1984,Hoover1985} was employed with a coupling constant of 1 ps. The pressure was maintained at 1 bar during both equilibration and production runs using the Berendsen barostat and Parrinello-Rahman barostat\cite{Parrinello1981}. For equilibration, a time constant of 5 ps and a compressibility of $4.5\times{10^{-5}} {\rm bar}^{-1}$ were employed. The Verlet cutoff scheme with a cutoff radius of 1.2 nm was utilized to construct the neighbor list in the all-atom MD simulations. Hydrogen bond lengths were constrained using the LINCS algorithm\cite{Hess1997}. A time step of 2 fs was employed for the all-atom molecular dynamics simulations. The particle mesh Ewald method\cite{Darden1993} was employed to calculate the electrostatic interactions.

\section{Results}

\subsection{Equilibration of aqueous $[{\rm ZnSO_{4}}]$ and $[{\rm MgSO_{4}}]$ mixtures: All-atom MD}
\begin{figure*}[hbtp]
\centering
\begin{subfigure}{1\columnwidth}
\centering
\caption{The dipole moment autocorrelation function vs time for ${\rm {[{ZnSO_{4}}]_{0M,1M,2M}+[{MgSO_{4}}]_{0M,1M,2M}}}$ mixtures}
\includegraphics[scale=1]{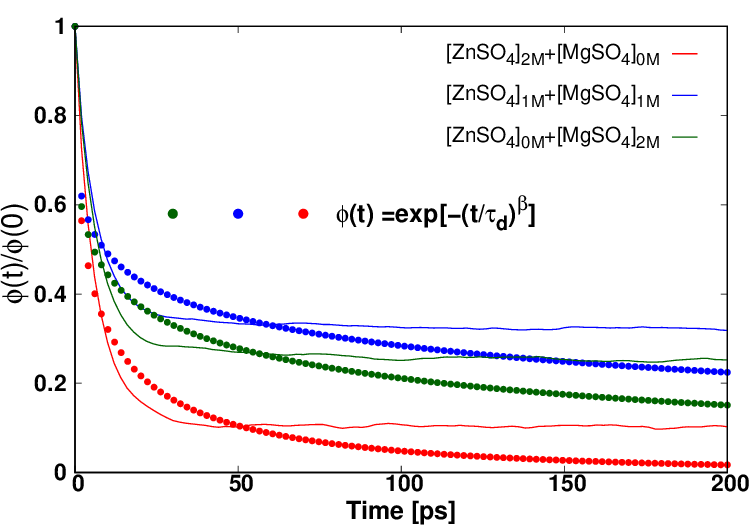}
\end{subfigure}
\vspace{10mm}

\begin{subfigure}{1\columnwidth}
 \centering
\caption{The average relaxation time ${{\tau}_{\rm d}}$.}
\begin{adjustbox}{width=0.60\columnwidth,center}
%\begin{tabular}{|c|wc{5cm}|wc{5cm}|}
\begin{tabular}{|c|c|c|}
\hline 
\multirow{2}{*}{{System Code}} & 
\multicolumn{2}{|c|}{
Relaxation time ${\tau}_{{\rm d}}$ [{\rm ps}]}  \\ 
\cline{2-3}
& $\beta=1$ & $\beta\ne 1$ \\
\hline
${\rm {[{ZnSO_{4}}]_{2M}+[{MgSO_{4}}]_{0M}}}$ & \quad ~~9.51$\pm$0.84 \quad & ~ 7.46$\pm$0.73 ~{($\beta=0.42)$}~~\quad\\ 
\hline 
${\rm {[{ZnSO_{4}}]_{1M}+[{MgSO_{4}}]_{1M}}}$   & \quad ~~79.00$\pm$5.66 \quad & ~ 39.56$\pm$2.79 ~{($\beta=0.24)$}~~\quad\\ 
\hline 
${\rm {[{ZnSO_{4}}]_{0M}+[{MgSO_{4}}]_{2M}}}$ & \quad ~~49.57$\pm$4.08 \quad & ~ 20.91$\pm$1.75 ~{($\beta=0.28)$}~~\quad\\ 
\hline 
\end{tabular}  
\end{adjustbox} 
%\caption{The average rotational relaxation time $\tau$ and stretching exponent $\beta$ of polymer chains for each melt system.}
\end{subfigure}
\caption{
The figure depicts the dipole moment autocorrelation function vs time for ${\rm {[{ZnSO_{4}}]_{2M}+[{MgSO_{4}}]_{0M}}}$, ${\rm {[{ZnSO_{4}}]_{1M}+[{MgSO_{4}}]_{1M}}}$, ${\rm {[{ZnSO_{4}}]_{0M}+[{MgSO_{4}}]_{2M}}}$ (a), (b) displays the relaxation time ${{\tau}_{\rm d}}$. The dipole moment autocorrelation function  ${\phi}(t)$ is fitted using the exponential function to determine the relaxation time ${{\tau}_{\rm d}}$.
}
\label{Figure_PHIt-TAUd}
\end{figure*}

In this section, we present the results obtained from our all-atom molecular dynamics simulations. We focus on investigating various properties of the ${\rm {[{ZnSO_{4}}]_{0M,1M,2M}+[{MgSO_{4}}]_{0M,1M,2M}}}$ mixtures, including dipole-dipole autocorrelation ${\phi}(t)$. These analyses are conducted for the mixture described in Table \ref{table:table_System}.
We have estimated the dipole moment autocorrelation function ${\phi}(t)$ for each mixture by using the following equation:
\begin{equation}
  \begin{minipage}[c]{0.80\linewidth}
   \centering
       $ {\phi}(t) = \displaystyle  \langle  {{\boldsymbol{\mu}}(t){\cdot}{{\boldsymbol{\mu}}}(0)}  \rangle $
\end{minipage}
\label{eqn:graphic_Phit}
\end{equation}

where the vectors ${\boldsymbol{\mu}}(0)$ and ${\boldsymbol{\mu}}(t)$ represent the dipole moment of the system
at time $t={\rm 0}$ and time $t$, respectively, while the angle bracket denotes an ensemble average.
We have illustrated the dipole moment autocorrelation function, denoted as ${\phi}(t)$ in Figure 3.
We determined the dipole relaxation time by fitting the dipole-dipole autocorrelation function to both a simple exponential function and a stretched exponential function \cite{Williams1970}, defined as follows:
\begin{equation}
  \begin{minipage}[c]{0.80\linewidth}
    \centering
          ${\phi}(t) =\displaystyle \exp\biggr[-\biggr({\frac{t}{\tau_{\rm d}}}\biggr)^{{\beta}}\biggr]$
  \end{minipage}
  \label{eqn:graphic_KWW}
\end{equation}
\noindent The dipole relaxation times, denoted as ${{\tau}_{\rm d}}$, and the associated stretching exponents, represented as $\beta$.
The dipole-dipole autocorrelation function ${\phi}(t)$ is also fitted using the Kohlrausch-Williams-Watts stretched exponential function to determine the dipole relaxation time ${{\tau}_{\rm d}}$ and the stretching exponent ${\beta}$ and the values are shown in Figure \ref{Figure_PHIt-TAUd}. 
From the comparison between Figure \ref{Figure_PHIt-TAUd}(b), we can see that the average relaxation times $\tau_{\rm d}$ obtained by using $\beta=1$ and $\beta < 1$ are similar, although the dipole relaxation times estimated by using stretched exponential functions are always respectively smaller than the ones by using simple exponential decay functions. 
%The relaxation behavior of ${\rm {[{ZnSO_{4}}]_{1M}+[{MgSO_{4}}]_{1M}}}$  substantially slower than that of ${\rm {[{ZnSO_{4}}]_{2M}+[{MgSO_{4}}]_{0M}}}$ and ${\rm {[{ZnSO_{4}}]_{0M}+[{MgSO_{4}}]_{2M}}}$. 
The relaxation behavior of the solution containing ${\rm 1M {[{ZnSO_{4}}]}}$ and ${\rm 1M {[{MgSO_{4}}]}}$ is significantly slower compared to the solutions with ${\rm 2M {[{ZnSO_{4}}]}}$ and no ${\rm {[{MgSO_{4}}]}}$, as well as the solutions with no ${\rm  {[{ZnSO_{4}}]}}$ and ${\rm 2M {[{MgSO_{4}}]}}$.

\subsection{Association and solvation structure of ${\rm {[{{Zn}^{2+}-SO_{4}}^{2-}]}}$ and ${\rm {[{{Mg}^{2+}-SO_{4}}^{2-}]}}$}
To explore the association of  ${\rm {[{{Zn}^{2+}-SO_{4}}^{2-}]}}$, ${\rm {[{{Mg}^{2+}-SO_{4}}^{2-}]}}$, ${\rm {[{{{Zn}^{2+}}-{{Zn}^{2+}}]}}}$,  ${\rm {[{{{Mg}^{2+}}-{{Mg}^{2+}}]}}}$ and,  ${\rm {[{{{Zn}^{2+}}-{{Mg}^{2+}}]}}}$, an analysis was conducted by calculating the radial distribution functions (RDFs) between the ${\rm {{{Zn}^{2+}}}}$, ${\rm {{{Mg}^{2+}}}}$, ${\rm {{SO_{4}}^{2-}}}$, and ${\rm {{H_{2}O}}}$. The ${\rm {[{{Zn}^{2+}-SO_{4}}^{2-}]}}$, ${\rm {[{{Mg}^{2+}-SO_{4}}^{2-}]}}$, ${\rm {[{{{Zn}^{2+}}-{{Zn}^{2+}}]}}}$,  ${\rm {[{{{Mg}^{2+}}-{{Mg}^{2+}}]}}}$ and,  ${\rm {[{{{Zn}^{2+}}-{{Mg}^{2+}}]}}}$ radial distribution function (RDFs) is defined as the ratio of the local density of the component site at distance r from the another component site and the bulk component density. The radial distribution functions ($g_{\alpha\beta}(r)$) are defined by the following equation:
\begin{equation}
  \begin{minipage}[c]{0.80\linewidth}
    \centering    
$g_{\alpha\beta}(r) ={{\displaystyle \langle \rho_\beta (r) \rangle_{{\rm local},{\alpha}}} \over {{\displaystyle \langle \rho_\beta (r_{\rm c}) \rangle_{\alpha}}}}$
  \end{minipage}
  \label{eqn:graphic_RDF}
\end{equation}

\begin{figure*}[hbtp]
\centering
\includegraphics[scale=1.0]{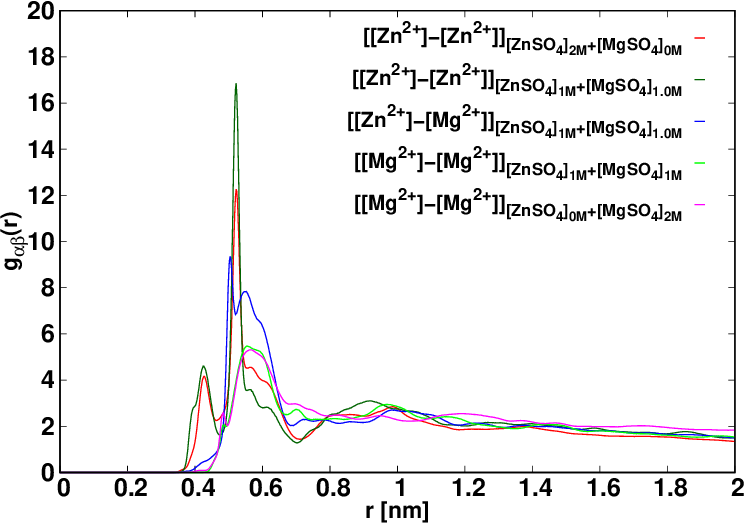}
\caption{
The radial distribution functions (RDFs) between ${\rm {[{{{Zn}^{2+}}-{{Zn}^{2+}}]}}}$, ${\rm {[{{{Zn}^{2+}}-{{Mg}^{2+}}]}}}$, and ${\rm {[{{{Mg}^{2+}}-{{Mg}^{2+}}]}}}$ ions.}
\label{fig:fig_RDF_ZNZN}
\end{figure*}

\noindent Herein, $\displaystyle \langle \rho_\beta (r) \rangle_{{\rm local},{\alpha}}$ represents the local mean particle density of ${\beta}$ particles in the vicinity of ${\alpha}$ particles, measured at a radial distance $r$. The denominator within the expression on the right-hand side of equation (\ref{eqn:graphic_RDF}), denoted as ${\displaystyle \langle \rho_\beta (r_{\rm c}) \rangle_{\alpha}}$, signifies the mean density of particles of type ${\beta}$ enclosed within a spherical volume of radius $r_{c}$, centered at the location of the ${\alpha}$ particle. The value of $r_{\rm c}$ being specifically established as half of the simulation box's dimensions.
Specifically, we estimated the RDFs of ${\rm {{{Zn}^{2+}}}}$ around ${\rm {{{Zn}^{2+}}}}$ (representing the RDF of zinc atom of ${\rm {{{Zn}^{2+}}}}$ ion around zinc atom of ${\rm {{{Zn}^{2+}}}}$ ion), ${\rm {{{Mg}^{2+}}}}$ around ${\rm {{{Mg}^{2+}}}}$ (representing the RDF of magnesium atom of ${\rm {{{Mg}^{2+}}}}$ ion around magnesium atom of ${\rm {{{Mg}^{2+}}}}$ ion), and ${\rm {{{Mg}^{2+}}}}$ around ${\rm {{{Zn}^{2+}}}}$ (representing the RDF of magnesium atom of ${\rm {{{Mg}^{2+}}}}$ ion around zinc atom of ${\rm {{{Zn}^{2+}}}}$ ion). Furthermore, we examined the RDFs of ${\rm {[{{Zn}^{2+}-SO_{4}}^{2-}]}}$ (representing the ${\rm {{SO_{4}}^{2-}}}$ ions around the ${\rm {{{Zn}^{2+}}}}$ ion) and ${\rm {[{{Mg}^{2+}-SO_{4}}^{2-}]}}$ (representing the ${\rm {[{SO_{4}}^{2-}}}$ ions around the ${\rm {[{{Mg}^{2+}}}}$ ion), ${\rm {[{{Zn}^{2+}-H_{2}O}]}}$ (representing the ${\rm {{H_{2}O}}}$ molecules around the ${\rm {{{Zn}^{2+}}}}$ ion),  ${\rm {[{{Mg}^{2+}-H_{2}O}]}}$ (representing the ${\rm {{H_{2}O}}}$ molecules around the ${\rm {{{Mg}^{2+}}}}$ ion).
Notably, two distinct peaks were observed at distances of 0.4 nm and 0.5 nm in the ${\rm {[{{{Zn}^{2+}}-{{Zn}^{2+}}]}}}$ RDFs (Figure \ref{fig:fig_RDF_ZNZN}), indicating a strong association between the ${\rm {[{{Zn}^{2+}}}}$ ions. In System III, the first RDF peak of ${\rm {[{{{Mg}^{2+}}-{{Mg}^{2+}}]}}}$ appears as a shoulder, while the second peak is broader. As the system transitions from System III to System II, the intensity of the first RDF peak increases, with a slight enhancement of the second peak. Additionally, two distinct peaks were observed in the ${\rm {[{{{Zn}^{2+}}-{{Mg}^{2+}}]}}}$ RDF for System II.

As the composition changes from ${\rm {[{ZnSO_{4}}]_{2M}+[{MgSO_{4}}]_{0M}}}$ to an equimolar mixture of 1M ${\rm {[{MgSO_{4}}]}}$ and 1M ${\rm {[{MgSO_{4}}]}}$ (${\rm {[{ZnSO_{4}}]_{1M}+[{MgSO_{4}}]_{1M}}}$) (Figure \ref{fig:fig_RDF_ZNZN}), the association between ${\rm {{{Zn}^{2+}}}}$ ions is significantly enhanced, while the association between ${\rm {{{Mg}^{2+}}}}$ ions is slightly enhanced. The association between ${\rm {[{{{Zn}^{2+}}-{{Mg}^{2+}}]}}}$ is substantially higher than that of ${\rm {[{{{Mg}^{2+}}-{{Mg}^{2+}}]}}}$ while smaller than ${\rm {[{{{Zn}^{2+}}-{{Zn}^{2+}}]}}}$.

\begin{figure*}[hbtp]
\centering
\includegraphics[scale=1.0]{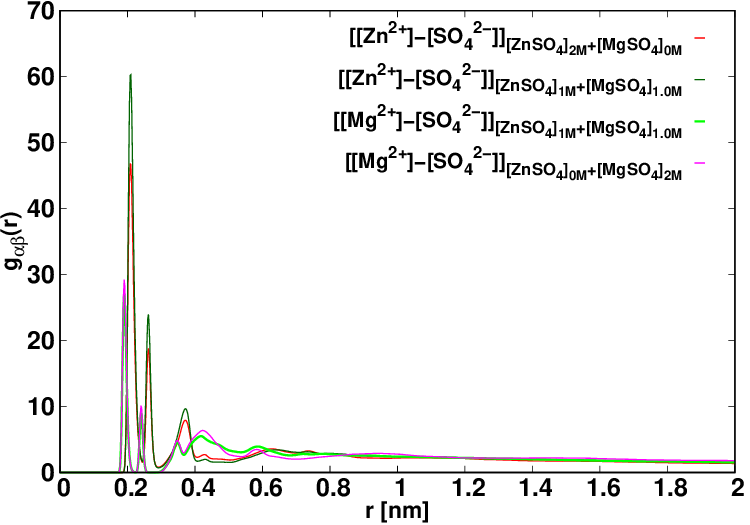}
\caption{
The radial distribution functions (RDFs) between ${\rm Zn^{2+}}-{\rm {SO_{4}}^{2-}}$ and ${\rm Mg^{2+}}-{\rm {SO_{4}}^{2-}}$ ion pairs.}
\label{fig:fig_RDF_ZNSO4ZNSO4}
\end{figure*}
 
In System I (${\rm {[{ZnSO_{4}}]_{2M}+[{MgSO_{4}}]_{0M}}}$), we observed three narrow RDF peaks for ${\rm {[{{Zn}^{2+}-SO_{4}}^{2-}]}}$. In System III (${\rm {[{ZnSO_{4}}]_{0M}+[{MgSO_{4}}]_{2M}}}$), two distinct RDF peaks for ${\rm {[{{Mg}^{2+}-SO_{4}}^{2-}]}}$ were identified. As the system transitions from System I to System II (${\rm {[{ZnSO_{4}}]_{1M}+[{MgSO_{4}}]_{1M}}}$), and from System III to System II, the intensities of the RDF peaks for${\rm {[{{Zn}^{2+}-SO_{4}}^{2-}]}}$ increase, while the peaks for ${\rm {[{{Mg}^{2+}-SO_{4}}^{2-}]}}$ show slight enhancement. This suggests that the structure of sulfate ions ${\rm SO_{4}}^{2-}$ surrounding ${\rm Zn}^{2+}$ is more ordered than that around ${ \rm MgSO_{4}}$. Furthermore, the sulfate ion structure around ${\rm Zn}^{2+}$ becomes more compact with the addition of ${ \rm MgSO_{4}}$.

\begin{figure*}[hbtp]
\centering
\includegraphics[scale=1.0]{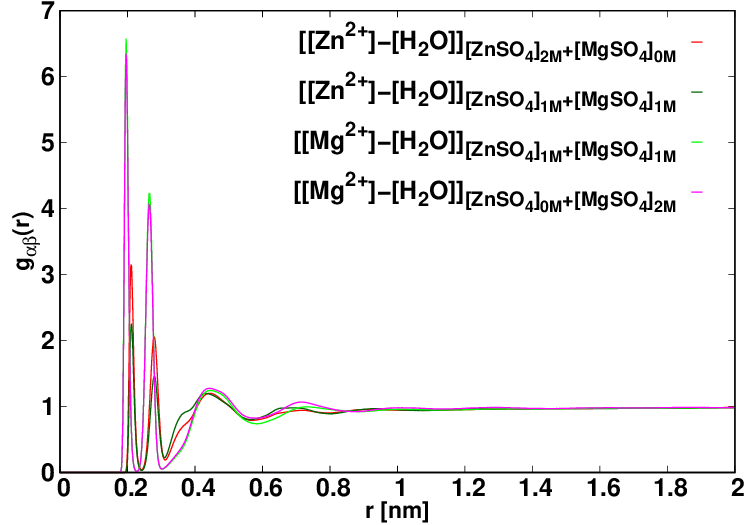}
\caption{
The radial distribution functions (RDFs) between ${\rm Zn^{2+}}-{\rm {H_{2}O}}$ and ${\rm Mg^{2+}}-{\rm {H_{2}O}}$ in different mixtures.}
\label{fig:fig_RDF_ZNwatZNwat}
\end{figure*}
   
\begin{figure*}[hbtp]
\centering
\begin{subfigure}[b]{1\linewidth}
\centering
\caption{ the running coordination numbers (RCNs), $n_{\alpha\beta}(r)$ of ${\rm Zn^{2+}}$ around ${\rm Zn^{2+}}$, ${\rm Mg^{2+}}$ around ${\rm Zn^{2+}}$, and ${\rm Mg^{2+}}$ around ${\rm Mg^{2+}}$ }
\centering
\includegraphics[scale=0.6]{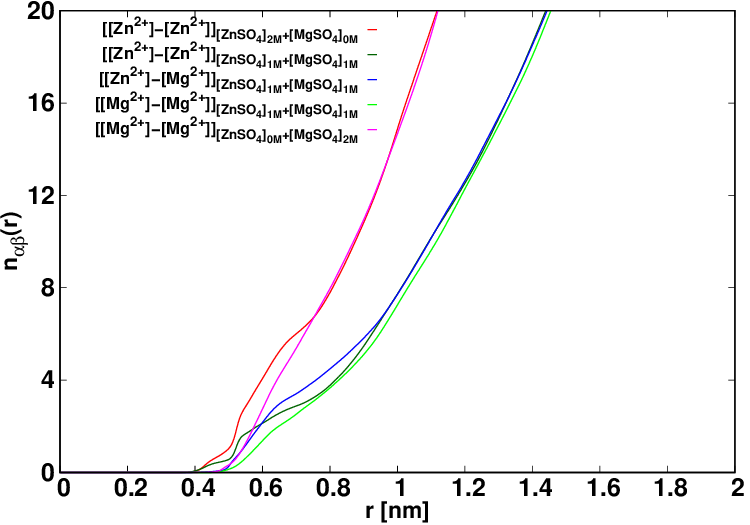}
\label{Figure_RCN-ZNZN-MGMG}
\end{subfigure}
\begin{subfigure}[b]{1\linewidth}
\centering
\caption{ The running coordination numbers (RCNs), $n_{\alpha\beta}(r)$ of  ${\rm {{SO_{4}}^{2-}}}$ around ${\rm Zn^{2+}}$,  ${\rm {{SO_{4}}^{2-}}}$ around ${\rm Mg^{2+}}$ }
\centering
\includegraphics[scale=0.6]{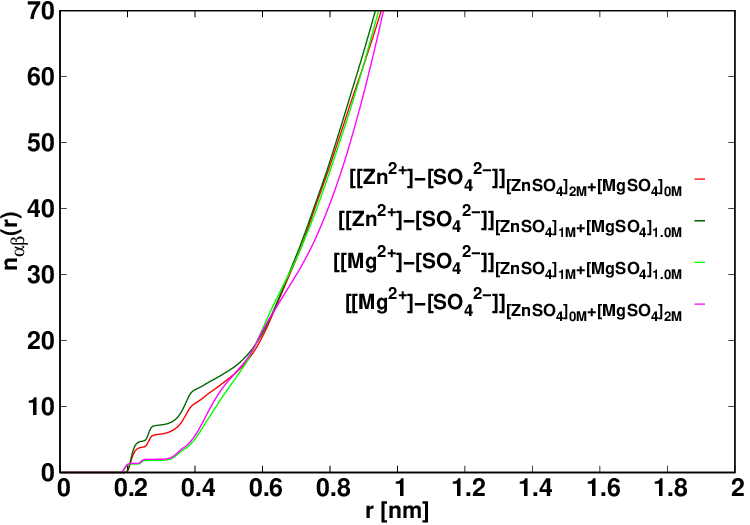}
\label{Figure_RCN-ZNSO4-MGSO4}
\end{subfigure}

\begin{subfigure}[b]{1\linewidth}
\centering
\caption{The running coordination numbers (RCNs), $n_{\alpha\beta}(r)$ of  ${\rm {{H_{2}O}}}$ around ${\rm Zn^{2+}}$,  ${\rm {{H_{2}O}}}$ around ${\rm Mg^{2+}}$ }
\centering
\includegraphics[scale=0.6]{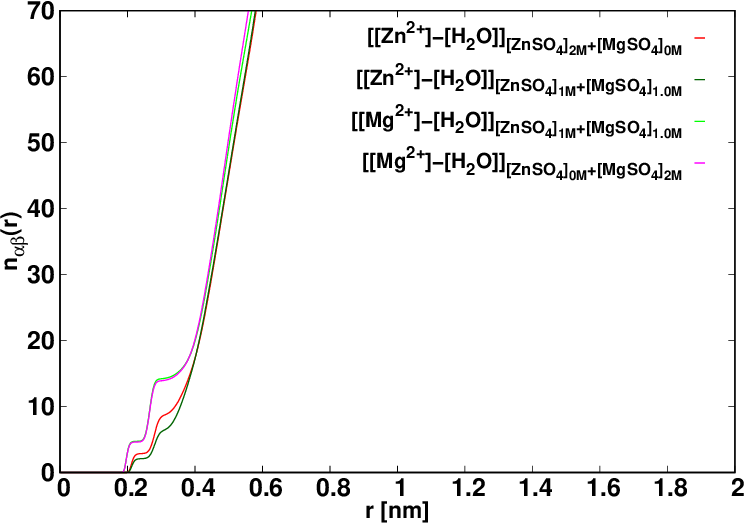}
\label{Figure_RCN-ZNWAT-MGWAT}
\end{subfigure}

\caption{
%\hl{
The figure caption illustrates the running coordination numbers (RCNs), $n_{\alpha\beta}(r)$ of ${\rm Zn^{2+}}$ around ${\rm Zn^{2+}}$, ${\rm Mg^{2+}}$ around ${\rm Zn^{2+}}$, and ${\rm Mg^{2+}}$ around ${\rm Mg^{2+}}$,${\rm {{SO_{4}}^{2-}}}$ around ${\rm Zn^{2+}}$,  ${\rm {{SO_{4}}^{2-}}}$ around ${\rm Mg^{2+}}$, ${\rm {{H_{2}O}}}$ around ${\rm Zn^{2+}}$,  ${\rm {{H_{2}O}}}$ around ${\rm Mg^{2+}}$  in the mixtures. 
The RCNs are calculated by considering ${\rm Zn^{2+}}$, ${\rm Mg^{2+}}$, ${\rm {{SO_{4}}^{2-}}}$, ${\rm {{H_{2}O}}}$ as ${\alpha}$ and ${\beta}$ particles in the expression $g_{\alpha\beta}(r)$. 
}
\label{Figure_RCN}
\end{figure*}

In System I (${\rm {[{ZnSO_{4}}]_{2M}+[{MgSO_{4}}]_{0M}}}$), two narrow RDF peaks were observed at 0.2 nm and 0.24 nm, along with two broader peaks at 0.42 nm and 0.7 nm for the  ${\rm Zn^{2+}}-{\rm {H_{2}O}}$ pair. Upon the addition of ${ \rm MgSO_{4}}$, the intensities of the narrow RDF peaks significantly decreased, while the intensities of the broader peaks showed a slight increase. This indicates that the water structure around ${Zn}^{2+}$ is initially highly compact, featuring four coordination shells. However, with the addition of ${ \rm MgSO_{4}}$, the water structure around ${\rm Zn}^{2+}$ becomes less compact.
In System III (${\rm {[{ZnSO_{4}}]_{0M}+[{MgSO_{4}}]_{2M}}}$), two narrow RDF peaks were detected at 0.19 nm and 0.22 nm, alongside two broader peaks at 0.43 nm and 0.72 nm for the ${\rm Mg^{2+}}-{\rm {H_{2}O}}$ pair. With the addition of ${ \rm ZnSO_{4}}$, the intensities of the narrow RDF peaks slightly increased, whereas those of the broader peaks slightly decreased. The presence of four distinct solvation shells around ${Mg}^{2+}$ confirms a highly ordered water structure. Upon the addition of ${ \rm ZnSO_{4}}$, the solvation shells around ${Mg}^{2+}$ become more compact.
%\hl{
We have investigated the solvation structure of ${\rm Zn}^{2+}$ around ${\rm Zn}^{2+}$, ${\rm Mg}^{2+}$ around ${\rm Zn}^{2+}$, ${\rm Mg}^{2+}$ around ${\rm Mg}^{2+}$, ${{{\rm SO_{4}}^{2-}}}$ ions around the ${\rm {{{Zn}^{2+}}}}$ ion, ${\rm {{SO_{4}}^{2-}}}$ ions around the ${\rm {{{\rm Mg}^{2+}}}}$ ion, the ${\rm {{H_{2}O}}}$ molecules around the ${\rm {{{Zn}^{2+}}}}$ ion, and the ${\rm {{H_{2}O}}}$ molecules around the ${\rm {{{Mg}^{2+}}}}$ ion. To achieve this, we employ Equation \ref{eqn:graphic_RCN} to compute the running coordination numbers (RCNs) delineating the spatial arrangement of ${\rm Zn}^{2+}$ around ${Zn}^{2+}$, ${Mg}^{2+}$ around ${\rm Mg}^{2+}$, ${\rm {{SO_{4}}^{2-}}}$ ions around the ${\rm {{{Zn}^{2+}}}}$ ion, ${\rm {{SO_{4}}^{2-}}}$ ions around the ${\rm {{{Mg}^{2+}}}}$ ion, the ${\rm {{H_{2}O}}}$ molecules around the ${\rm {{{Zn}^{2+}}}}$ ion, and the ${\rm {{H_{2}O}}}$ molecules around the ${\rm {{{Mg}^{2+}}}}$ ion. The coordination number is defined as:
%}
\begin{equation}
  \begin{minipage}[c]{0.90\linewidth}
    \centering
${\displaystyle n_{\alpha\beta}}=\displaystyle { 4\pi{\rho}_{\beta}}{\int_{r_{\rm 1}}^{r_{\rm 2}}{r^{2}g_{\alpha\beta}(r)dr}}$
  \end{minipage}
  \label{eqn:graphic_RCN}
\end{equation}
%\hl{
In this context, $n_{\alpha\beta}$ denotes the number of  
 type $\beta$ atoms surrounding species $\alpha$, confined within a radial shell extending from $r_{1}$ to $r_{2}$. Here, ${\rho}_{\beta}$  represents the number density of $\beta$ in the system, while $g_{\alpha\beta}(r)$ stands for the radial distribution function. The latter provides the ratio of the local density of $\beta$ around $\alpha$ to the bulk density of $\beta$. For the specific calculation of the first solvation shell coordination number, $r_{1}$ is set to zero, signifying the immediate vicinity of the species $\alpha$, and $r_{2}$ corresponds to the position of the first minimum observed in the radial distribution function. This method captures the nuanced spatial arrangement of atoms surrounding a central species.
The running coordination numbers (RCNs), $n_{\alpha\beta}$, are depicted in Figure \ref{Figure_RCN}.
%}

Figure \ref{Figure_RCN-ZNZN-MGMG} demonstrates that the population of ${\rm Zn^{2+}}$ ions within the first, second, and third coordination shells surrounding other ${\rm Zn^{2+}}$ ions significantly decreases upon the addition of ${\rm MgSO_{4}}$. Similarly, the population of ${\rm Mg^{2+}}$ ions around other ${\rm Mg^{2+}}$ ions in the first and second coordination shells decreases when ${\rm ZnSO_{4}}$ is added. In the 1M ${\rm ZnSO_{4}}$ and 1M  ${\rm MgSO_{4}}$ mixture, the number of ${\rm Zn^{2+}}$ ions in the coordination shells around ${\rm Zn^{2+}}$ ions is considerably higher than the number of ${\rm Mg^{2+}}$ ions around ${\rm Mg^{2+}}$ ions. 
%%%%%%%%%%%%%%%%%%%%%%%%%%%%%%%%%%%%%%%%%%%%%%%%%%%%%%%%%%%
Figure \ref{Figure_RCN-ZNSO4-MGSO4} clearly shows that the population of ${\rm {{SO_{4}}^{2-}}}$ ions around ${\rm Zn^{2+}}$ ions is significantly larger than that around ${\rm Mg^{2+}}$ ions in the first, second, and third coordination shells. The number of ${\rm {{SO_{4}}^{2-}}}$ ions around ${\rm Zn^{2+}}$ ions in these coordination shells increases substantially from system-I to system-III, while the number around ${\rm Mg^{2+}}$ ions slightly decreases from system-III to system-II.
Analysis of Figure \ref{Figure_RCN-ZNWAT-MGWAT} indicates that the number of water molecules around ${\rm Zn^{2+}}$ ions in the first, second, and third coordination shells in the 2M ${\rm ZnSO_{4}}$ ${+}$ 0M ${\rm MgSO_{4}}$ mixture is significantly lower than the number around ${\rm Mg^{2+}}$ ions in the 0M ${\rm ZnSO_{4}}$ ${+}$ 2M ${\rm MgSO_{4}}$ mixture. As we transition from system-I to system-II, the number of water molecules around ${\rm Zn^{2+}}$ ions decreases substantially. Conversely, the population of water molecules around ${\rm Mg^{2+}}$ ions in the first, second, and third coordination shells slightly increases from system-III to system-II.
%%%%%%%%%%%%%%%%%%%%%%%%%%%%%%%%%%%%%%%%%%%%%%%%%%%%%%%%%%%
\newpage
The potentials of mean force (PMFs) are widely utilized to investigate the stability of clusters, as demonstrated in various studies.\cite{Chandler1987,Siddique2016,Dixit2017,Dixit2022,Dixit2015,Dixit2016,Sarkar2015,Dixit2016a,Meti2019,Dixit2021,Hajari2022,Ghosh2016,Meti2018,Kumar2019,Chatterjee2013,Dixit2013,Siddique2013,Jain2014,Dixit2020,Dixit2024polymerAufinal,Dixit2023bio,Dixit2024macro,Dixit2024JPCB}
 Accordingly, in this study, we have computed the potentials of mean force (PMFs) between ${\rm {[{{Zn}^{2+}-SO_{4}}^{2-}]}}$, and ${\rm {[{{Mg}^{2+}-SO_{4}}^{2-}]}}$, employing the following equation: 
\begin{equation}
  \begin{minipage}[c]{0.80\linewidth}
    \centering
        $W(r)=-RT {\rm log} (g(r))$  
  \end{minipage}
  \label{eqn:graphic_PMF}
\end{equation} 

\begin{figure*}[hbtp]
\centering
\includegraphics[scale=1.0]{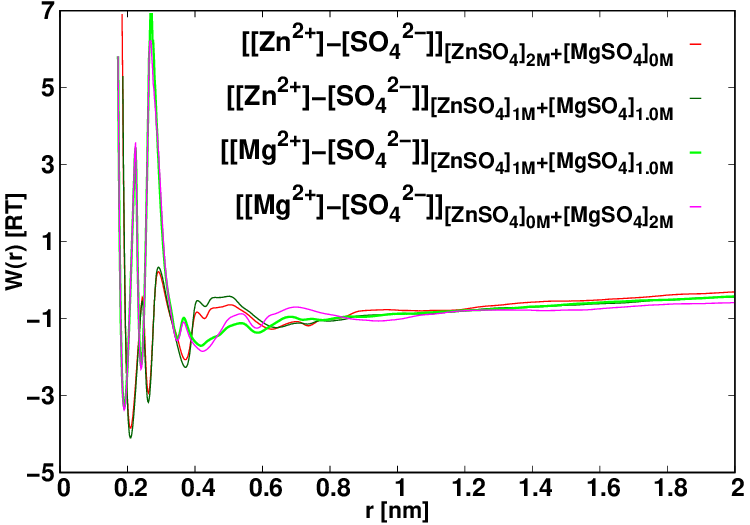}
\caption{
We calculate the Potentials of Mean Force (PMF) $W(r)$ using the equation $W(r)=-RT{\rm log}(g(r))$. Here, $R$ represents the molar gas constant in units of ${\rm kJmol^{-1}}/{\rm K}$, $T$ denotes the system temperature, and $g(r)$ corresponds to the radial distribution function between the ions. The magnitude of error bars in each PMF profile is less than 0.5 ${R}T$.
}
\label{Figure_PMF}
\end{figure*}

\noindent Here, $R$ signifies the  molar gas constant (in ${\rm kJmol^{-1}}/{\rm K}$), $T$ denotes the system's temperature, and $g(r)$ represents the radial distribution function between ${\rm {[{{Zn}^{2+}-SO_{4}}^{2-}]}}$, and ${\rm {[{{Mg}^{2+}-SO_{4}}^{2-}]}}$. 
Figure \ref{Figure_PMF} illustrates the potentials of mean force (PMFs) between the ${\rm {[{{Zn}^{2+}-SO_{4}}^{2-}]}}$, and ${\rm {[{{Mg}^{2+}-SO_{4}}^{2-}]}}$ as a function of their distance. 

The potentials of mean force (PMF), W(r), for the ion pairs ${\rm {[{{Zn}^{2+}-SO_{4}}^{2-}]}}$, and ${\rm {[{{Mg}^{2+}-SO_{4}}^{2-}]}}$ are depicted in Figure \ref{Figure_PMF}. These PMFs were calculated using equation \ref{eqn:graphic_PMF}. For the ${\rm {[{{Zn}^{2+}-SO_{4}}^{2-}]}}$ pair, the PMF reveals the presence of a distinct contact ion pair (CIP) at approximately 0.20 nm, followed by a solvent-assisted ion pair (SAIP) near 0.23 nm, and a solvent-shared ion pair (SShIP) at around 0.39 nm. Comparing systems I and II, we observe a slight increase in the stability of CIP, SAIP, and SShIP structures.
In the case of the ${\rm {[{{Mg}^{2+}-SO_{4}}^{2-}]}}$ pair, a CIP emerges at approximately 0.19 nm, followed by an SAIP at 0.22 nm, a SShIP at 0.32 nm, and a solvent-separated ion pair (SSIP) near 0.41 nm. The transition from system III to system II does not notably affect the stability of the CIP, SAIP, or SShIP; however, there is a reduction in SAIP stability. Notably, the energy barriers between the CIP, SAIP, and SShIP states for the ${\rm {[{{Mg}^{2+}-SO_{4}}^{2-}]}}$ pair are considerably higher than those observed for the ${\rm {[{{Zn}^{2+}-SO_{4}}^{2-}]}}$ pair.
\newpage
\subsection{Preferential binding coefficients}
 The preferential binding coefficient (${\gamma}$) is defined as follows:\cite{Record1995}
 \begin{equation}
  \begin{minipage}[c]{0.80\linewidth}
    \centering
        $ \displaystyle {\gamma} ={\displaystyle \left\langle {{n_{\rm SO_{4}}(r)} -\frac{{N_{\rm SO_{4}}}-n_{\rm SO_{4}}(r)}{{N_{w}}-n_{w}(r)}{{n_{w}(r)}}} \right\rangle}$ 
  \end{minipage}
  \label{eqn:graphic_Pref}
\end{equation}
In the context where ${n_{\rm SO_{4}^{2-}}(r)}$ denotes the count of ${\rm SO_{4}^{2-}}$ ions and ${n_{w}(r)}$ represents the count of water molecules situated at a radial distance $r$ from the center of mass of either the ${\rm Zn^{2+}}$ or ${\rm Mg^{2+}}$ ion, ${N_{\rm {SO_{4}}^{2-}}}$ signifies the total count of ${\rm SO_{4}^{2-}}$ ions, and ${N_{W}}$ signifies the total count of water molecules within the system. The sign of ${\gamma}$ assumes significance in elucidating ion behavior, where a positive value signifies a preference for ion binding with ${\rm SO_{4}^{2-}}$ ions, whereas a negative value indicates a propensity for the ion to favor proximity to water molecules.

\begin{figure*}[hbtp]
\centering
\includegraphics[scale=1.0]{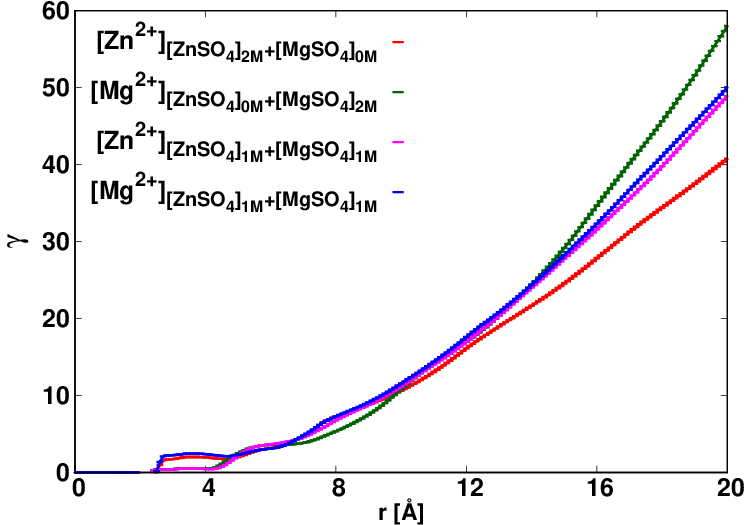}
\caption{
Preferential binding coeﬃcients (${\gamma}$) of ${\rm Zn^{2+}}$ and ${\rm Mg^{2+}}$ ions in ${\rm {[{ZnSO_{4}}]_{2M}+[{MgSO_{4}}]_{0M}}}$, ${\rm {[{ZnSO_{4}}]_{1M}+[{MgSO_{4}}]_{1M}}}$, and ${\rm {[{ZnSO_{4}}]_{0M}+[{MgSO_{4}}]_{2M}}}$ mixtures.
}
\label{Figure_Prefbinding}
\end{figure*}

\begin{table}[htbp]
\centering
\begin{adjustbox}{width=0.8\columnwidth,center}
\begin{tabular}{|l|c|c|}
\hline
\multirow{2}{*}{Systems}  & \multicolumn{2}{|c|} {Preferential binding coeﬃcients (${\gamma}$)}   \\
\cline{2-3}
 & {${\rm Zn^{2+}}$}  &{${\rm Mg^{2+}}$}  \\
\hline
${\rm {[{ZnSO_{4}}]_{2M}+[{MgSO_{4}}]_{0M}}}$ & 34.24$\pm$0.04  & -  \\
\hline
${\rm {[{ZnSO_{4}}]_{1M}+[{MgSO_{4}}]_{1M}}}$  & 39.76$\pm$0.03  &  40.89$\pm$0.08 \\
\hline
${\rm {[{ZnSO_{4}}]_{0M}+[{MgSO_{4}}]_{1M}}}$  &  - & 45.63$\pm$0.06  \\
\hline
\end{tabular}
\end{adjustbox}
\caption{Preferential binding coeﬃcients (${\gamma}$) of ${\rm Zn^{2+}}$ and ${\rm Mg^{2+}}$ ions in ${\rm {[{ZnSO_{4}}]_{2M}+[{MgSO_{4}}]_{0M}}}$, ${\rm {[{ZnSO_{4}}]_{1M}+[{MgSO_{4}}]_{1M}}}$, and ${\rm {[{ZnSO_{4}}]_{0M}+[{MgSO_{4}}]_{2M}}}$ mixtures.}
\label{table:table_Pref}
\end{table}
%%%%%%%%%%%%%%%%%%%%%%%%%%%%%%%%%%%%%%%%%%%%%%%%%%%%%%%%%%%%%
When ${\rm SO_{4}^{2-}}$ interacts with an ion, it replaces the water molecules surrounding the ${\rm Zn^{2+}}$ or ${\rm Mg^{2+}}$ ions. A positive preferential binding coefficient, ${\gamma}$, indicates a significant accumulation of ${\rm SO_{4}^{2-}}$ ions near the cation, while a negative ${\gamma}$ reflects a depletion of ${\rm SO_{4}^{2-}}$ ions from the vicinity of the cation. The preferential binding coefficients for ${\rm Zn^{2+}}$ and ${\rm Mg^{2+}}$  in the mixtures ${\rm {[{ZnSO_{4}}]_{2M}+[{MgSO_{4}}]_{0M}}}$, ${\rm {[{ZnSO_{4}}]_{1M}+[{MgSO_{4}}]_{1M}}}$, and ${\rm {[{ZnSO_{4}}]_{0M}+[{MgSO_{4}}]_{2M}}}$ were calculated using Equation \ref{eqn:graphic_Pref}, with the results shown in Table \ref{table:table_Pref}. Notably, the ${\gamma}$ values for ${\rm SO_{4}^{2-}}$ are consistently positive, implying that both ${\rm Zn^{2+}}$ and ${\rm Mg^{2+}}$ are preferentially solvated by ${\rm SO_{4}^{2-}}$ in all mixture compositions.
As the system transitions from ${\rm {[{ZnSO_{4}}]_{2M}+[{MgSO_{4}}]_{0M}}}$ to ${\rm {[{ZnSO_{4}}]_{1M}+[{MgSO_{4}}]_{1M}}}$, the preferential binding coefficient ${\gamma}$ for ${\rm Zn^{2+}}$ increases, indicating a stronger affinity of ${\rm Zn^{2+}}$ for ${\rm SO_{4}^{2-}}$ in the equimolar mixture. In contrast, as the system shifts from ${\rm {[{ZnSO_{4}}]_{0M}+[{MgSO_{4}}]_{2M}}}$ to ${\rm {[{ZnSO_{4}}]_{1M}+[{MgSO_{4}}]_{1M}}}$, the ${\gamma}$ value for ${\rm Mg^{2+}}$ decreases, suggesting a diminished affinity of ${\rm Mg^{2+}}$ for ${\rm SO_{4}^{2-}}$ in the equimolar system.
%%%%%%%%%%%%%%%%%%%%%%%%%%%%%%%%%%%%%%%%%%%%%%%%%%%%%%%%%%%%%5
\section{Discussion and Conclusions}
\indent In this investigation, we have undertaken a comprehensive study to elucidate the association and solvation structure of ${\rm {Zn}^{2+}-{SO_{4}^{2-}}}$ and ${\rm {Mg}^{2+}-{SO_{4}^{2-}}}$ ion pairs in ${\rm {[{ZnSO_{4}}]_{2M}+[{MgSO_{4}}]_{0M}}}$, ${\rm {[{ZnSO_{4}}]_{1M}+[{MgSO_{4}}]_{1M}}}$, and ${\rm {[{ZnSO_{4}}]_{0M}+[{MgSO_{4}}]_{2M}}}$ mixtures. 
To gain in-depth insights, we employed all-atom molecular dynamics (MD) simulations.
 We investigated the dynamics of ${\rm {[{ZnSO_{4}}]_{2M}+[{MgSO_{4}}]_{0M}}}$, ${\rm {[{ZnSO_{4}}]_{1M}+[{MgSO_{4}}]_{1M}}}$, and ${\rm {[{ZnSO_{4}}]_{0M}+[{MgSO_{4}}]_{2M}}}$ mixtures by analyzing the dipole-dipole autocorrelation function ${\phi}(t)$ and the dipole relaxation time ${{\tau}_{\rm d}}$.
Additionally, we explored the spatial arrangements of ${\rm Zn^{2+}}$ around ${\rm Zn^{2+}}$, ${\rm Mg^{2+}}$ around ${\rm Mg^{2+}}$,${\rm Mg^{2+}}$ around ${\rm Zn^{2+}}$, ${\rm SO_{4}^{2-}}$ around ${\rm Zn^{2+}}$, ${\rm SO_{4}^{2-}}$ around ${\rm Mg^{2+}}$, and ${\rm H_{2}O}$ around ${\rm Zn^{2+}}$, ${\rm H_{2}O}$ around ${\rm Mg^{2+}}$, using radial distribution functions (RDFs) and Running Coordination numbers.
We also estimated the potentials of mean force between ${\rm {Zn}^{2+}-{SO_{4}^{2-}}}$ and ${\rm {Mg}^{2+}-{SO_{4}^{2-}}}$ ion pairs in ${\rm {[{ZnSO_{4}}]_{2M}+[{MgSO_{4}}]_{0M}}}$, ${\rm {[{ZnSO_{4}}]_{1M}+[{MgSO_{4}}]_{1M}}}$, and ${\rm {[{ZnSO_{4}}]_{0M}+[{MgSO_{4}}]_{2M}}}$ mixtures.
Furthermore, we examined the preferential binding coefficient of ${\rm Zn^{2+}}$  and ${\rm Mg^{2+}}$ in ${\rm {[{ZnSO_{4}}]_{2M}+[{MgSO_{4}}]_{0M}}}$, ${\rm {[{ZnSO_{4}}]_{1M}+[{MgSO_{4}}]_{1M}}}$, and ${\rm {[{ZnSO_{4}}]_{0M}+[{MgSO_{4}}]_{2M}}}$ mixtures.
%%%%%%%%%%%%%%%%%%%%%%%%%%%%%%%%%%%%%%%%%%%%%%%%%%%%%%%%%%%%%%%%%%%%%%%%%%%%%
The relaxation behavior of the solution containing equimolar concentrations of ${\rm 1M {[{ZnSO_{4}}]}}$ and ${\rm 1M {[{MgSO_{4}}]}}$ exhibits significantly slower dynamics compared to solutions with ${\rm 2M {[{ZnSO_{4}}]}}$ in the absence of ${\rm {[{MgSO_{4}}]}}$ and solutions with ${\rm 2M {[{MgSO_{4}}]}}$ without ${\rm {[{ZnSO_{4}}]}}$ (Figure \ref{Figure_PHIt-TAUd}). This observation suggests that the presence of both ${\rm Zn^{2+}}$ and ${\rm Mg^{2+}}$ ions in the equimolar mixture leads to more complex and hindered relaxation processes, possibly due to competitive interactions and solvation dynamics involving ${\rm SO_{4}^{2-}}$ ions.
As the system transitions from a solution containing ${\rm 2M {{MgSO_{4}}]}}$ and no ${\rm {[{MgSO_{4}}]}}$ (${\rm {[{ZnSO_{4}}]_{2M}+[{MgSO_{4}}]_{0M}}}$)  to an equimolar mixture of ${\rm 1M {[{ZnSO_{4}}]}}$ and ${\rm 1M {[{MgSO_{4}}]}}$, a significant reduction in the association between ${\rm {{{{Zn}^{2+}}}}}$ ions is observed (Figure \ref{fig:fig_RDF_ZNZN}). This indicates a disruption in the self-association of ${\rm {{{{Zn}^{2+}}}}}$ due to the presence of ${\rm {[{{{Mg}^{2+}}}}}$. In contrast, the interaction between ${\rm {{{{Mg}^{2+}}}}}$ ions shows a slight increase, suggesting a modest enhancement in the association of ${\rm {{{{Mg}^{2+}}}}}$ ions as the concentration of ${\rm {[{ZnSO_{4}}]}}$ rises. These trends reflect the differential impacts of ion pairing and competition between ${\rm {[{{{Zn}^{2+}}}}}$ and ${\rm {[{{{Mg}^{2+}}}}}$ in mixed salt systems, with ${\rm {{{{Mg}^{2+}}}}}$ ions exhibiting a greater propensity for association in mixed salt solutions.
In System I (${\rm {[{ZnSO_{4}}]_{2M}+[{MgSO_{4}}]_{0M}}}$), the radial distribution function (RDF) for the ${\rm {[{{Zn}^{2+}-SO_{4}}^{2-}]}}$ ion pairs exhibits three distinct and sharp peaks, indicating a well-structured solvation environment around ${\rm Zn^{2+}}$ (Figure \ref{fig:fig_RDF_ZNSO4ZNSO4}). Conversely, System III (${\rm {[{ZnSO_{4}}]_{0M}+[{MgSO_{4}}]_{2M}}}$), reveals two well-defined RDF peaks for the ${\rm {[{{Mg}^{2+}-SO_{4}}^{2-}]}}$ ion pairs, suggesting a different level of structural organization around ${\rm Mg^{2+}}$.
As the system transitions to System II (${\rm {[{ZnSO_{4}}]_{1M}+[{MgSO_{4}}]_{1M}}}$), an increase in the intensity of the RDF peaks for ${\rm {[{{Zn}^{2+}-SO_{4}}^{2-}]}}$ is observed, indicating a more pronounced ordering of ${\rm {SO_{4}}^{2-}}$ ions around ${\rm Zn^{2+}}$. Similarly, the RDF peaks for the ${\rm {[{{Mg}^{2+}-SO_{4}}^{2-}]}}$ ion pairs show a slight enhancement, though less significant than that for ${\rm Zn^{2+}}$. These trends suggest that ${\rm {SO_{4}}^{2-}}$ ions exhibit a higher degree of structural organization around ${\rm Zn^{2+}}$ compared to ${\rm Mg^{2+}}$. Additionally, the increasing peak intensities in the presence of ${ \rm MgSO_{4}}$ imply that the solvation structure around ${\rm Zn^{2+}}$ becomes more compact, highlighting the stronger interaction between ${\rm Zn^{2+}}$ and ${\rm {SO_{4}}^{2-}}$ in mixed salt environments.
In conclusion, the results indicate that ${\rm Zn^{2+}}$ ions maintain a more robust and ordered coordination structure with ${\rm {SO_{4}}^{2-}}$ ions than ${\rm Mg^{2+}}$, and that the addition of ${ \rm MgSO_{4}}$ enhances the compactness and organization of ${\rm {SO_{4}}^{2-}}$ ions around ${\rm Zn^{2+}}$.
In System I (${\rm {[{ZnSO_{4}}]_{2M}+[{MgSO_{4}}]_{0M}}}$), the radial distribution function (RDF) for the${\rm Zn^{2+}}-{\rm {H_{2}O}}$ reveals two sharp peaks at 0.20 nm and 0.24 nm, along with two broader peaks at 0.42 nm and 0.70 nm (Figure \ref{fig:fig_RDF_ZNwatZNwat}). These observations indicate a highly structured and compact solvation environment around ${Zn}^{2+}$, characterized by four well-defined coordination shells. Upon the addition of ${ \rm MgSO_{4}}$, the intensities of the sharper peaks decrease markedly, while the broader peaks show a slight increase in intensity. This suggests that the addition of ${ \rm MgSO_{4}}$ leads to a disruption of the tightly packed solvation structure around ${Zn}^{2+}$, resulting in a more diffuse water arrangement.
Similarly, in System III (${\rm {[{ZnSO_{4}}]_{0M}+[{MgSO_{4}}]_{2M}}}$),  the ${\rm Mg^{2+}}-{\rm {H_{2}O}}$ RDFs exhibit distinct peaks at 0.19 nm and 0.22 nm, along with broader peaks at 0.43 nm and 0.72 nm, reflecting the presence of four coordination shells around ${Mg}^{2+}$. Upon the introduction of ${ \rm ZnSO_{4}}$, a slight enhancement in the intensities of the narrower peaks is observed, while the broader peaks experience a mild reduction. This indicates that the addition of ${ \rm ZnSO_{4}}$ further compacts the solvation structure around ${Mg}^{2+}$, leading to a more ordered water arrangement.
These findings suggest that in both systems, the solvation structures of water around ${Zn}^{2+}$ and ${Mg}^{2+}$ ions exhibit high degrees of order, with well-defined coordination shells. However, the introduction of the competing salt induces noticeable changes in the solvation structures of water, either disrupting or enhancing the compactness of the water molecules around the respective cations.
Figure \ref{Figure_RCN-ZNZN-MGMG} illustrates a significant decrease in the population of ${\rm Zn^{2+}}$  ions in the first, second, and third coordination shells surrounding other ${\rm Zn^{2+}}$  ions upon the addition of ${\rm MgSO_{4}}$. Similarly, the introduction of ${\rm ZnSO_{4}}$ results in a decrease in the population of ${\rm Mg^{2+}}$  ions in the first and second coordination shells surrounding other ${\rm Mg^{2+}}$  ions. In the equimolar mixture of 1M ${\rm ZnSO_{4}}$ and 1M ${\rm MgSO_{4}}$, the coordination environment is characterized by a considerably higher number of ${\rm Zn^{2+}}$  ions surrounding other ${\rm Zn^{2+}}$  ions compared to the number of ${\rm Mg^{2+}}$  ions surrounding other ${\rm Mg^{2+}}$  ions.
Further analysis, as shown in Figure \ref{Figure_RCN-ZNSO4-MGSO4}, reveals that the population of ${\rm {{SO_{4}}^{2-}}}$ ions around ${\rm Zn^{2+}}$  ions is significantly larger than that surrounding ${\rm Mg^{2+}}$  ions across all coordination shells. Notably, the number of ${\rm {{SO_{4}}^{2-}}}$ ions associated with ${\rm Zn^{2+}}$  ions in these shells increases substantially from System I to System II, while the number associated with ${\rm Mg^{2+}}$  ions exhibits a slight decrease from System III to System II.
In addition, Figure \ref{Figure_RCN-ZNWAT-MGWAT} indicates that the population of water molecules around ${\rm Zn^{2+}}$  ions in the first, second, and third coordination shells within the 2M ${\rm ZnSO_{4}}$ +0M${\rm MgSO_{4}}$ mixture is significantly lower than that around ${\rm Mg^{2+}}$  ions in the 0M ${\rm ZnSO_{4}}$ +2M${\rm MgSO_{4}}$ mixture. As the system transitions from System I to System II, the number of water molecules surrounding ${\rm Zn^{2+}}$  ions decreases substantially, whereas the population of water molecules around ${\rm Mg^{2+}}$  ions in the first, second, and third coordination shells shows a slight increase from System III to System II.
These observations highlight the distinct differences in solvation dynamics between ${\rm Zn^{2+}}$  and ${\rm Mg^{2+}}$  ions, suggesting that the solvation environment becomes less compact around ${\rm Zn^{2+}}$  in the presence of ${\rm MgSO_{4}}$, while the hydration shell around ${\rm Mg^{2+}}$  may exhibit increased stabilization with higher ${\rm ZnSO_{4}}$ concentrations.
The potentials of mean force (PMF), denoted as W(r), for the ion pairs ${\rm {[{{Zn}^{2+}-SO_{4}}^{2-}]}}$, and ${\rm {[{{Mg}^{2+}-SO_{4}}^{2-}]}}$  are illustrated in Figure \ref{Figure_PMF}. For the ${\rm {[{{Zn}^{2+}-SO_{4}}^{2-}]}}$ pair, the PMF analysis indicates the formation of a distinct contact ion pair (CIP), succeeded by a solvent-assisted ion pair (SAIP) and a solvent-shared ion pair (SShIP). A comparative analysis of systems I and II reveals a slight enhancement in the stability of CIP, SAIP, and SShIP structures.
Conversely, for the ${\rm {[{{Mg}^{2+}-SO_{4}}^{2-}]}}$ pair, the PMF identifies a CIP, followed by an SAIP, a SShIP, and a solvent-separated ion pair (SSIP). Transitioning from system III to system II does not significantly alter the stability of the CIP, SAIP, or SShIP states; however, there is a notable decrease in the stability of the SAIP. Importantly, the energy barriers separating the CIP, SAIP, and SShIP configurations for the ${\rm {[{{Mg}^{2+}-SO_{4}}^{2-}]}}$ pair are considerably higher than those observed for the ${\rm {[{{Zn}^{2+}-SO_{4}}^{2-}]}}$ pair.
These findings suggest that while both ion pairs exhibit similar structural motifs, the interactions involving  ${\rm Mg^{2+}}$   and ${\rm {{SO_{4}}^{2-}}}$ are characterized by higher energetic barriers, indicating a less favorable stabilization compared to their ${\rm Zn^{2+}}$ counterparts. This differential behavior in stability and energy barriers may have significant implications for understanding the solvation dynamics and coordination chemistry of these ions in solution.
The preferential binding coefficients (${\gamma}$) for ${\rm Zn^{2+}}$ and ${\rm Mg^{2+}}$ in the mixtures ${\rm {[{ZnSO_{4}}]_{2M}+[{MgSO_{4}}]_{0M}}}$, ${\rm {[{ZnSO_{4}}]_{1M}+[{MgSO_{4}}]_{1M}}}$, and ${\rm {[{ZnSO_{4}}]_{0M}+[{MgSO_{4}}]_{2M}}}$ were determined (Figure \ref{Figure_Prefbinding}), with the resulting values summarized in Table \ref{table:table_Pref}. A noteworthy observation is that the ${\gamma}$ values for ${\rm SO_{4}^{2-}}$  consistently exhibit positive values across all mixture compositions. This indicates that both ${\rm Zn^{2+}}$ and ${\rm Mg^{2+}}$ are preferentially solvated by ${\rm SO_{4}^{2-}}$.
As the system transitions from ${\rm {[{ZnSO_{4}}]_{2M}+[{MgSO_{4}}]_{0M}}}$ to ${\rm {[{ZnSO_{4}}]_{1M}+[{MgSO_{4}}]_{1M}}}$,  there is an observable increase in the preferential binding coefficient ${\gamma}$ for ${\rm Zn^{2+}}$ , suggesting a strengthened affinity of ${\rm Zn^{2+}}$ for ${\rm SO_{4}^{2-}}$ in the equimolar mixture. Conversely, when the system shifts from ${\rm {[{ZnSO_{4}}]_{0M}+[{MgSO_{4}}]_{2M}}}$ to ${\rm {[{ZnSO_{4}}]_{1M}+[{MgSO_{4}}]_{1M}}}$, the ${\gamma}$ value for ${\rm Mg^{2+}}$ exhibits a decrease, indicating a reduced affinity of ${\rm Mg^{2+}}$ for ${\rm SO_{4}^{2-}}$ in the equimolar system.
These findings suggest that the preferential solvation behavior of these divalent cations is significantly influenced by their concentrations within the mixed solutions. The increased affinity of ${\rm Zn^{2+}}$ for ${\rm SO_{4}^{2-}}$ compared to ${\rm Mg^{2+}}$may have implications for understanding the dynamics of ion interactions in various chemical environments and could inform future studies on solvation and ion association in electrolyte systems.
\section*{Data Availability Statement}
In this study, we performed extensive all-atom molecular dynamics (MD) simulations using the GROMACS software suite, specifically versions 5.1.4 and 2019.2, which were critical to our computational analysis. The respective software can be accessed at the following links:\url{https://manual.gromacs.org/documentation/5.1.4/download.html} and \url{https://manual.gromacs.org/documentation/2019.2/download.html}. The MD simulations were executed using input scripts tailored for our systems, along with a custom-developed Fortran code designed for the calculation of the potential of mean force (PMF) from radial distribution functions. This code is publicly available on GitHub at \url{https://github.com/mayankmoni/PMF-from-gr}.
Furthermore, the preferential binding coefficients for the systems under investigation were determined through computational analysis using another Fortran code, which can be accessed at \url{https://github.com/mayankmoni/Preferential-Binding-Coef}. These resources are made openly available to promote reproducibility and facilitate further investigation. Additional research data supporting the findings of this work can be provided upon request.
\section*{Acknowledgement}
MD thanks Japan Science and Technology Agency (JST) for funding (the grant number JPMJCR2091). Our appreciation extends to the "Joint Usage/Research Center for Interdisciplinary Large-scale Information Infrastructures" and the "High-Performance Computing Infrastructure" in Japan (Project ID Nos.: jh210017-MDH, jh220054, jh230061, and jh240063) for their essential computational resources. The utilization of the Wisteria/BDEC-01 system at the Information Technology Center, University of Tokyo, has significantly contributed to the success of our research.
\section*{Funding sources}
This research did not receive any specific grant from funding agencies in the public, commercial, or not-for-profit sectors.
\section*{Author Contributions}
Mayank Dixit: Conceptualization, Methodology, Software , Data curation, Writing- Original draft preparation, Visualization, Investigation. Bhalachandra Laxmanrao Tembe: Supervision.: Timir Hajari  Writing- Reviewing and Editing
%\end{acknowledgement}

%%%%%%%%%%%%%%%%%%%%%%%%%%%%%%%%%%%%%%%%%%%%%%%%%%%%%%%%%%%%%%%%%%%%%%%%%%%%%
%\begin{thebibliography}

%\bibliographystyle{elsarticle-harv} 
\bibliographystyle{elsarticle-num} 
\bibliography{library-mgso4znso4}

%\end{thebibliography}
\end{document}